\documentstyle[12pt]{article}
\begin{document}

\title{Continuous variable quantum cloning via the Cavity QED with trapped ions}
\author{XuBo Zou, K. Pahlke and W. Mathis  \\
\\Institute TET, University of Hannover,\\
Appelstr. 9A, 30167 Hannover, Germany }
\date{}

\maketitle

\begin{abstract}
{\normalsize We present a feasible scheme to use trapped ions
Cavity QED system to implement optimal $1\rightarrow N$ cloning
machine of coherent state. In present scheme, as the ouput of the
clone machine, the copies of the cavity mode emerge at the
vibrational modes and cavity modes is acted as the ancilla mode in
the implementation of the cloning. We further show that such
system can be used to generated multimodes quantum state, which
can realize optimum symmetric $1\rightarrow N$ telecloning of
coherent state of distant cavity.}

\end{abstract}

It is not possible to construct a device that produce an exact
copy of an arbitrary quantum system\cite{ww}. This result can be
seen as a consequence of the linearity of quantum mechanics. Since
the seminal paper of Buzek and Hillery\cite{bm}, quantum cloning
has been extensively studied theoretically for discrete quantum
variable, such as quantum bits\cite{nd} or d-level
system\cite{wbc}. The experimental scheme for cloning qubit has
been proposed via stimulated emission\cite{sk}. Recently various
concepts of the quantum information processing have been extended
to the domain of continuous quantum variables. The established
continuous variables protocols comprise of the quantum state
teleportation\cite{bo}, quantum error correction\cite{slb} and
quantum cryptography\cite{rh}. Recently. the quantum cloning of
the continuous variable system has also been considered in a state
dependent context. In Ref\cite{cr}, the duplication of the
coherent state was investigated. The optimality of this continuous
variable cloning transformation was proved in Ref\cite{cs}. More
recently, $N\rightarrow M$ cloning of the coherent state has been
proposed based on the
linear amplifer and beam splitter\cite{aaa}.
On the other hand, telecloning of coherent state has also been considered\cite{pvl}\\
Recent advance in laser cooling and ion trapping have opened
prospect in quantum information processing using trapped ions.
Starting with the ion trap computer proposed by Cirac and
Zoller\cite{cz}, a number of schemes have been proposed for
quantum computation and entangling internal ion
states\cite{bb,bbb}. The approach with a bichromatic
field\cite{bb} led to the successful entanglement of four trapped
ions\cite{cas}. All these proposals are mostly concerned with the
determinstic manipulation of the internal states of the ions. In
Ref\cite{ah}. the scheme has been proposed to enable motional
quantum state to be coupled to propagating light field via
interaction in Cavity QED. Such scheme can be used to generate EPR
state in the position and momentum of distantly seperated trapped
ions\cite{ash} and delocalized mesoscopic vibrational
state\cite{al}. In addition, the ponderomotive force scheme is
also proposed to entangle atomic motion of the two trapped ion and
therefore entangle two continuous variable (infinite dimensional)
system based on trapped ion Cavity QED system\cite{sh}. Hence it
is expected that trapped ions system can be used to implement
quantum computation and quantum communication of continuous
variable. In this paper, we present a scheme to implement the
$1\rightarrow N$ Gaussian continuous variable quantum cloning
machine via trapped ion Cavity QED. We further show that such
system can be used to generate multimodes quantum state, which can
realize optimum symmetric $1\rightarrow N$
telecloning of coherent state\cite{pvl}. \\
The basic setup we use to couple atomic motion to light was
considered by A. S. Parkins and E. Larabal\cite{al}. This setup
consists of N two-level atoms confined in microtraps located in an
optical cavity. The kind of situation one might imagine is like
that realized in the experiment of Ye et al\cite{jy}, in which
atoms were trapped inside a microscopic optical cavity using a
far-off-resonance dipole force trap. The atomic transition of
frequency $\omega_0$ is coupled to a single mode of the cavity
field of frequency $\omega_{cav}$. A single coupling laser field
of frequency $\omega_{A}$ is incident on the atoms from the side
of the cavity. We shall consider the large detuning of the
internal atomic transition from laser field so that atomic
spontaneous emission can be neglected and the internal atomic
dynamics can be adiabatically eliminated. The Hamiltonian for the
system can be written as ( assuming a single microtrap frequency
$\nu_{x}$ and tight confinement along ttransverse direction as
well)
$$
H=\sum_j\nu_x(b_j^{\dagger}b_j+\frac{1}{2})+\delta_{cA}a^{\dagger}a
+\sum_j\frac{g_0^2}{\Delta_{oA}}\sin^2(kx_j+\theta_j)a^{\dagger}a
$$
$$
+\sum_j\frac{g_0\epsilon_A}{\Delta_{oA}}\sin(kx_j+\theta_j)(e^{-i\varphi_A}a^{\dagger}+e^{i\varphi_A}a)
\eqno{(1)}
$$
where $a$ and $b_j$ $(j=1\cdots N)$ are annihilation operators for
the cavity mode and vibrational modes of N trapped ions. The
single photon atom cavity dipole coupling strength is given by
$g_0$. $k$ is the wave vector of the cavity field,
$kx_j=\eta_j(b_j^{\dagger}+b_j)$, $\eta_j$ is the corresponding
Lamb-Dicke parameter. The detuning $\delta_{cA}$ and $\Delta_{oA}$
are given by $\delta_{cA}=\omega_{cav}-\omega_{A}$ and
$\Delta_{oA}=\omega_{o}-\omega_{A}$. $\theta_j$ define the
position of the center of the jth microtrap relative to a node of
the cavity QED field. $\varphi_A$ and $ \epsilon_A$ are the phase
and amplitude of the laser field. \\We now choose the detuning
between the cavity and laser field to be $\delta_{cA}=\nu_x$. By
using the rotating wave approximation, we obtain the effective
interaction Hamiltonian between the cavity and vibrational mode
(in the Lamb-Dicke limit)
$$
H_1=\Omega_1e^{i\varphi_A}a(b_1^{\dagger} +\cdots
b_N^{\dagger})+\Omega_1e^{-i\varphi_A}a^{\dagger}(b_1 +\cdots b_N)
\eqno{(2)}
$$
If we tune the laser field to satisfy $\delta_{cA}=-\nu_x$, we
obtain, after discarding the rapidly oscillating terms, effective
interaction
$$
H_1=\Omega_2e^{i\varphi_A}a(b_1 +\cdots
b_N)+\Omega_1e^{-i\varphi_A}a^{\dagger}(b_1^{\dagger} +\cdots
b_N^{\dagger}) \eqno{(3)}
$$
Here $\Omega_1$ and $\Omega_2$ are effective interaction
couplings. We assume that the ions-laser interaction strength is
same for all ions.\\In what follows, we show trapped ions Cavity
QED system can be used to implement $1 \rightarrow N$ Gaussian
cloning machine of cavity field state. Let us begin with $1
\rightarrow 2$ cloner. Suppose that the cavity field state is
initially in the coherent state $|\xi>$ and two vibrational modes
$b_1$, $b_2$ are in the vaccum states, i.e. the total three modes
state can be written as $|\Psi(0)>=|\xi>_a|0>_{b_1}|0>_{b_2}$. In
the following, we see at the output of the present clone machine,
two duplicate of the input state $|\xi>_a$ emerge at the two
vibrational modes $b_1$, $b_2$, the cavity field mode is acted as
the ancillia mode in the implement of cloner. The unitary cloning
transformation performed by clone machine can be realized by
trapped ion cavity QED. We first apply the laser pulse with
interaction (1) to derive the trapped ions and choose the
amplitude (or interaction time $t_1$) and phase of laser field in
such a way that the condition $\varphi_A=\frac{\pi}{2}$ and
$\Omega_1t_1=\frac{\pi}{2\sqrt{2}}$ are fulfilled, we obtain state
at time $t_1$
$$
|\Psi(t_1)>=\exp(-iH_1t_1)|\xi>_a|0>_{b_1}|\xi>_{b_2}=|0>_a|\frac{\xi}{\sqrt{2}}>_{b_1}|\frac{\xi}{\sqrt{2}}>_{b_2}
\eqno{(4)}
$$
We then tune the laser pulse and derive the ions with ion-laser
interaction (3). If the amplitude (or interaction time $t_2$) and
phase of laser field is chosen to satisfy
$\varphi_A=\frac{\pi}{2}$ and
$\cosh(\sqrt{2}\Omega_2t_2)=\sqrt{2}$, we obtain state of the
system
$$
|\Psi(t_1+t_2)>=\exp\left( v[a(b_1+ b_2)-a^{\dagger}(b_1^{\dagger}
+b_2^{\dagger})]\right
)|0>_a|\frac{\xi}{\sqrt{2}}>_{b_1}|\frac{\xi}{\sqrt{2}}>_{b_2}
\eqno{(5)}
$$
Where $v=\frac{arc\cosh\sqrt{2}}{\sqrt{2}}$. The Q function are
same for two cloned state $\rho_{b_1}$ and $\rho_{b_2}$
corresponding to the state $|\Psi(t_1+t_2)>$
$$
Q(\alpha)=\frac{2}{3\pi}\exp(-\frac{2}{3}|\alpha-\xi|^2)
\eqno{(6)}
$$
and both copies have the same fidelity $F=\frac{2}{3}$, which is
exactly the upper bound of the fidelity for $1 \rightarrow 2$
Gaussian clone machine, so that our cloning transformation is
optimal.symmetric $1\rightarrow N$
telecloning of coherent state\cite{pvl}. For initial state \\
The scheme can be extended to design symmetric cloning machine
which prepare N ( $N\geq3$) identical approximate copies from a
single cavity field. Such clone machine consist of the N trapped
ions in optical Cavity. The N available copies of the cavity field
subjected to cloning are stored in vibrational modes $b_1, \cdots,
b_N$. We consider the situation that the cavity field is in the
coherent state $|\xi>$ and N vibrational modes are in the vaccum
states, $|\Psi(0)>=|\xi>_a|0>_{b_1}\cdots|0>_{b_N}$. If the
interaction (2) and (3) are applied in success to initial state
and durations $t_1$ and $t_2$ of the respective interaction (or
the corresponding interaction strength) are chosen to be
$\Omega_1t_1=\frac{\pi}{2\sqrt{N}}$ and
$\cosh(\sqrt{N}\Omega_2t_2)=\sqrt{N}$ and phase of laser field is
set to be $\varphi_A=\frac{\pi}{2}$, the system evolve to
$$
|\Psi(t_1+t_2)>=\exp\left(
\frac{arc\cosh\sqrt{N}}{\sqrt{N}}[a(b_1+\cdots+
b_N)-a^{\dagger}(b_1^{\dagger}+\cdots+b_N^{\dagger})]\right
)
$$
$$
\times
|0>_a|\frac{\xi}{\sqrt{N}}>_{b_1}\cdots|\frac{\xi}{\sqrt{N}}>_{b_N}
\eqno{(7)}
$$
The Q function of the N output modes can be expressed as
$$
Q(\alpha)=\frac{N}{(2N-1)\pi}\exp(-\frac{N}{2N-1}|\alpha-\xi|^2)
\eqno{(8)}
$$
and corresponding fidelity is $F=\frac{N}{2N-1}$, which is exactly
the upper bound of the fidelity for $1 \rightarrow N$ cloner
derived in Ref. Hence, machine is a optimal $1 \rightarrow N$
cloner of coherent state\\
In what follows, we will show that such system can also be used to
generate multimode quantum state, which can realize the optimum
symmetric $1\rightarrow N$ telecloning of coherent state. For
initial state$|\Psi(0)>=|0>_a|0>_{b_1}\cdots|0>_{b_N}$, we apply
the laser pulse with interaction $H_2$ and choose amplitude (or
interaction time $t_2$) and phase of laser field to satisfy
$$\varphi_A=\frac{\pi}{2}~~~~\Omega_2t_2=\frac{r}{\sqrt{N}}
\eqno{(9)}$$ where
$$
e^{2r}=\frac{\sqrt{N}+1}{\sqrt{N}-1} \eqno{(10)}
$$ and we obtain state of the system
$$
|\Psi(t_2)>=\exp(\frac{r}{\sqrt{N}}[a(b_1+\cdots+
b_N)-a^{\dagger}(b_1^{\dagger}+\cdots+b_N^{\dagger})])
|0>_a|0>_{b_1}\cdots|0>_{b_N} \eqno{(11)}
$$
whose wigner function is
$$W(x_a,x_{b_1},\cdots,x_{b_N},p_a,p_{b_1},\cdots,p_{b_N})=(\frac{2}{\pi})^{N+1}\exp[-e^{-2r}(x_a+\frac{1}{\sqrt{N}}
\sum_{i=b_1}^{b_N}x_i)^2
$$
$$
-e^{2r}(p_a+\frac{1}{\sqrt{N}} \sum_{i=b_1}^{b_N}p_i)^2
-e^{2r}(x_a-\frac{1}{\sqrt{N}}
\sum_{i=b_1}^{b_N}x_i)^2
$$
$$
-e^{-2r}(p_a-\frac{1}{\sqrt{N}} \sum_{i=b_1}^{b_N}p_i)^2
-\frac{1}{N}\sum_{i,j=b_1}^{b_N}[(x_i-x_j)^2+(p_i-p_j)^2] ]
\eqno{(12)}$$ which is exactly state discussed in Ref \cite{pvl}.
Similar to proposal for teleportation of the wave function of a
massive particleRef \cite{ass}, we can also use state to realize
the optimum symmetric $1\rightarrow N$ telecloning of coherent
state of distant cavity.\\
In summary, we have proposed a feasible scheme to use trapped ions
Cavity QED system to implement optimal $1\rightarrow N$ cloning
machine of coherent state. In present scheme, as the ouput of the
clone machine, the copies of the cavity mode emerge at the
vibrational modes and cavity modes is acted as the ancilla mode in
the implementation of the cloning. We further show that such
system can be used to generated multimodes quantum state, which
can realize optimum symmetric $1\rightarrow N$ telecloning of
coherent state of distant cavity.

\end{document}